\documentclass[12pt]{article}
\usepackage{amsmath,amsfonts, amssymb,graphicx,geometry,authblk,color,soul,comment,hyperref, algorithm, algpseudocode} 
\DeclareMathOperator*{\argmin}{arg\,min}
\DeclareMathOperator*{\argmax}{arg\,max}

\title{Constitutive relations from images}
\author{ Adeline Wihardja and Kaushik Bhattacharya}
\affil{California Institute of Technology}
\date{}

\begin{document}
\maketitle

\begin{center}
{\it Dedicated to Richard D. James whose work exemplifies the interplay between theory and experiment, and fidelity to both experimental observations and mathematical rigor.}
\end{center}

\begin{abstract}
Constitutive relations close the balance laws of continuum mechanics and serves as the surrogate for a material in the design and engineering process.  The problem of obtaining the constitutive relations is an indirect inverse problem where both the relation and the quantities that define the relation have to be inferred from experimental observations.  The advent of full-field observation techniques promises a new ability of learning constitutive relations in realistic operational conditions.  However, this is done in two steps, first obtaining deformations from the images, and then obtaining the constitutive relation from deformations and forces.  This leads to a variety of difficulties.  In this paper, we propose a novel approach that enables us to obtain constitutive relation directly from the raw data consisting of images and force measurements.

\end{abstract}

\section{Introduction}

The design of structural, aerospace, protection and other thermomechanical engineering systems requires a constitutive relation that describes the material properties \cite{gurtin_1970,billington_1981}, and is empirically determined.  The essential challenge is that we cannot measure these constitutive relations directly; in fact we cannot even measure the quantities like stress, heat flux, energy density and state variables that describe the constitutive relations.  

The traditional approach relies on uniform states (e.g., uniaxial tension) or universal solutions (e.g. torsion)  \cite{billington_1981}.  Each test provides limited information, and they require repetition.  This has been addressed with automation and high-throughput approaches \cite{miracle_2024}.  However, this is not always possible since some experiments (e.g., plate impact)  need sterile conditions with precise alignment to obtain the desired state \cite{meyers}.  In any case, all of these still sample  idealized states of strain.  These limit the fidelity of the resulting models when we use them to model complex situations.  We propose to develop methods that can provide high fidelity constitutive relations by probing complex domains and complex states by using full field techniques like digital image correlation (DIC) \cite{sutton2009image,hild} and thermal imaging.

DIC takes a series of images of a decorated surface and compares them to infer the deformation  \cite{sutton2009image,hild}.  It has emerged as a method of choice due to the ease of using it with the widespread availability of cameras and robust commercial (e.g., \cite{vic}) and open source software (e.g., \cite{aldic}).  The full-field information it provides also gives useful insight.  It has been extended beyond optical imaging (e.g., \cite{kammers_digital_2013} using scanning electron microscopy images), to digital volume correlation \cite{buljac_digital_2018,yang_augmented_2020}, and tomography \cite{mehdikhani}.
The deformation is then used to obtain constitutive relations.

This current practice of splitting the inverse problem into two inverse problems, first image to deformation, and second deformation and force to constitutive relation  poses a number of challenges.  First, these inverse problems may be ill-posed and have to be regularized.  When they are done separately, the regularization of one problem can limit the information available to the other problem.  Indeed,  DIC as a purely kinematic problem of inferring deformation is ill-posed.  In practice, it is either regularized using filters or constrained by finite dimensional (local or global) ansatz.  However, finite dimensional approximations do not necessarily converge, and regularization limits the efficacy when there are cracks, shear bands and shocks.  Further, we image a fixed region of space that may correspond to different parts of the body in different snapshots.  So, image registration limits us to small displacements.  One can stitch together multiple  snapshots, but errors may be compounded, and the method is inaccessible to dynamic events where one has very few snapshots.

Second, even when a reasonable deformation is available, the second inverse problem is still a difficult problem since we cannot directly measure stress, but have to infer it from the total force acting on a part of the boundary.  So it is still common to use simple configurations, though the full-field inversion is a subject of active research, and we refer the reader to \cite{akerson_learning_2024} for a recent literature survey.

Third, we may have a mismatch of sensitivities.  Even if the overall sensitivity is reasonable, the splitting into two problems leading to situations where the first inversion has low sensitivity, but the second has high sensitivity.  This leads to significant errors.  

Finally, there are phenomena where the instrument and the experiment are intimately coupled, rendering a sequential inversion impossible or dependent on unverifiable ansatz.  For example, Lawlor {\it et al.} \cite{lawlor_2024} recently introduced a speckle pattern in the interior of a transparent specimen and were able to observe the interaction between a shock wave and a pore.  However, due to the photoelastic effect that changes the refractive index under stress, the image is heavily distorted in the vicinity of the shock as well in the wake of the shock.  

In this paper, we propose an alternate approach where we study the integrated inverse problem of inferring constitutive relations directly from raw data in the form of images of speckle patterns and overall forces.  We formulate this as an optimization problem, and solve it using a gradient-based approach.  We focus on finite elasticity and rubber, and demonstrate the method using both synthetic and experimental data.

The study of the mechanical properties of rubber goes back to seminal works of Treloar \cite{treloar1944}, Rivlin and Saunders \cite{rivlinsaunders51} and Gent and Rivlin \cite{gent1952experiments}.  Assuming incompressibility, one can fully characterize the material by applying two principals independently, and this motivated the biaxial tension test \cite{rivlinsaunders51} that is widely used \cite{abeyaratne,tokumoto}.  These and other approaches \cite{kawabata1981experimental,becker1967phenomenological,blatzko} seek to create a uniform state.  Others use universal solutions \cite{ericksen} like torsion \cite{treloar1944} and inflation \cite{gent1952experiments}.   More recent efforts study complex deformations in two and three dimensions exploiting advances in optical microscopy \cite{parsons2004experimental,grytten2009,rouxhild12} and x-ray computed tomography \cite{angkur}.  We consider optimal imaging of complex two dimensional domains.

We provide the formulation in Section \ref{sec:form}.  We demonstrate the approach using synthetic data in Section \ref{sec:syn}, and experimental observations in Section \ref{sec:exp}.  We conclude in Section \ref{sec:conc}

\section{Formulation} \label{sec:form}

We assume that the material is hyperelastic, and governed by a stored energy per unit reference volume $W$.  We assume a parametrized form, so that $W=W(F;P)$ where $F$ is the deformation gradient, and $P\in {\mathbb R}^P$ is a set of parameters.  Our goal is to find the parameters from an experiment.

%
\subsection{Image to constitutive relation}

We consider a body occupying the region $\Omega \subset {\mathbb R}^n$ in the natural reference configuration.  We fix part of the boundary $\partial_0 \Omega$, and apply a time dependent Dirichlet boundary condition $y(x,t) = \bar{y}(x,t)$ on another part $\partial_y \Omega$.   The remainder of the boundary is traction-free.  At each instant $t$, the body is at equilibrium, and the deformation $y:\Omega \to {\mathbb R}^n$ satisfies 
\begin{equation} 
\nabla \cdot W_F (\nabla y; P) = 0 \text{ on } \Omega
\end{equation}
subject to the boundary conditions where $W_F = \partial W/\partial F$.  Equivalently, 
\begin{align} \label{eq:gov}
  -  \int_\Omega \left(W_F (\nabla y; P) \cdot \nabla \varphi \right) \, d\Omega = 0  \quad \quad \forall \ \varphi \in \mathcal{U} = \{u = 0 \text{ on } \partial_0 \Omega \cup \partial_y \Omega \}.
\end{align}

We decorate a part of the traction-free boundary of the body with a speckle (or other distinctive) pattern, and image a certain region of space ${\mathcal R}$ that contains a part of the traction-free part of the boundary to obtain images $g(y,t)$ (we typically obtain a series of snapshots).  We also measure the total reaction force on the part $\partial_y \Omega$.   

Our goal is to obtain $P$ from the observations $\{g(y,t), f(t)\}$, by matching them with the corresponding quantities computed with a model with parameters $P$.  We formulate this as an optimization problem
\begin{equation}
    P = \argmin \mathcal{O}(P),
\end{equation} 
where the objective is 
\begin{equation} \label{eq:obj}
    \mathcal{O}= \int_0^t w(t) \left( \int_{\mathcal{S}} \left| g_0(x) - g(y(x,t),t) \right|^2 \, dA + 
    \alpha \left| \int_{\partial_y\Omega}W_F(\nabla y(x,t);P) \hat{n} \, dA - f_\text{exp}(t) \right|^2 \right) \, dt,
\end{equation}
and $y(x,t)$ satisfies the governing equation (\ref{eq:gov}), $w(t), \alpha$ are weights, and $g_0(x) = g(x,0)$.

We solve this by a gradient descent, and therefore seek to calculate the sensitivity of the objective with respect to the parameters.  Doing so directly requires us to calculate $y_P$, the sensitivity of the solution to the governing equation (\ref{eq:gov}) to the parameters;  this is difficult.  Therefore, we use the adjoint equation \cite{oberai_solution_2003,bonnet_inverse_2005}.   To derive this equation, add the left hand side of (\ref{eq:gov}) to our objective, and then differentiate with respect to P.    We obtain,
\begin{equation}
\begin{aligned}
\frac{d  \mathcal{O}}{dP}  = & -2 \int_0^t \left( \int_{\mathcal{S}} w(t) (g(x,0) - g(y(x,t),t)) g_y y_P \, dA \right.\\
& + 2 \alpha  w(t) \left( \int_{\partial_y\Omega}W_F \hat{n}  \, dA - f_\text{exp}(t) \right) \cdot
\left( \int_{\partial_y\Omega} \left( W_{FF} \nabla y_P + W_{FP}  \right) \hat{n}\, dA \right) \\
&\left.+ \int_\Omega \left( W_{FF} \nabla y_p + W_{FP} \right) \cdot \nabla \varphi \, dx \right) \, dt
\end{aligned}
\end{equation}
where $W_F, W_{FF}, W_{FP}$ are all evaluated at $(\nabla y (x,t); P)$.  This still contains the problematic quantity $y_P$.  However, this expression is true for all $\varphi \in \mathcal{U}$ (cf.\ (\ref{eq:gov}).   We now make a special choice such that we eliminate $y_P$ from above.  Specifically, let $\varphi$ satisfy the {\it adjoint equation}
\begin{equation} \label{eq:adj}
 \int_\Omega \nabla \psi \cdot W_{FF} \nabla \varphi \, dx + \int_{\mathcal S} \mu(x,t) \psi \, dA + 
 \int_{\partial_y \Omega}  \nabla \psi  \cdot W_{FF} (\lambda(t) \otimes \hat{n}) = 0 \, dA \quad \quad \forall \ \psi \in \mathcal{U}
\end{equation}
where we have used the fact that $W_{FF}$ has major symmetry and 
\begin{equation}
\mu(x,t) = -2 w(t) (g_0(x) - g(y(x,t),t)) g_y, \quad \lambda(t) = 2\alpha w(t)\left( \int_{\partial_y\Omega}W_F \hat{n} \, dA - f_\text{exp}(t) \right).
\end{equation}
Then, the sensitivity of the objective with respect to the parameter is
\begin{equation} \label{eq:sen}
\frac{d  \mathcal{O}}{dP}  =   \int_0^t  \left(  \int_\Omega W_{FP}(\nabla y (x,t);P) \cdot \nabla \varphi \, dx
+ \lambda(t)  \cdot \int_{\partial_y \Omega} W_{FP} \, \hat{n} \, dA \right) \, dt.
\end{equation}

\subsection{Numerical method}

\begin{algorithm}[t]
\caption{Constitutive relations from images \label{alg:algo}}
\textbf{Input:} Reference image $g_0$, deformed images $g$, parametrized model  \\
\textbf{Output:} Displacement $y$
\begin{algorithmic}
    \State \textbf{Step 1:} Initiate guess of parameters $P$;
    \State \textbf{Step 2:} Pick region of interest (${\mathcal S}$) to perform correlation; 
    \State \textbf{Step 3:} Determine locations of each pixel in each finite element within ${\mathcal S}$; 
    \State \textbf{Step 4:} Precompute spatial gradients of the image $\nabla g$; 
    \While{$\|d\mathcal{O}/dP\| \, \textbf{or} \, \|\mathcal{O}_{k}-\mathcal{O}_{k-1}\| > \varepsilon$}
         \State \textbf{Step 5:} Solve forward problem for displacement $y_{k}(x,t)$;
          \For{\textbf{each}\textit{ pixel in each finite element in ${\mathcal S}$}}
             \State \textbf{Step 6:} Compute $g(y(x,t),t)$, $g_y$ from $y_{k}(x,t)$;
            \EndFor
        \State \textbf{Step 7:} Solve inverse problem for adjoint variable $\delta y_{k}(x,t)$;
        \State \textbf{Step 8:} Compute objective $\mathcal{O}(y_{k},P_k)$;
        \State \textbf{Step 9:} Compute sensitivity $d\mathcal{O}(y_k,P_k)/dP$ for each $P$;
        \State \textbf{Step 10:} Update parameters $P_{k+1} = P_k + \delta P$ with MMA;
    \EndWhile
\end{algorithmic}
\end{algorithm}

We start with an initial guess for the parameters, and then solve the forward and inverse problem using a finite element method, use the sensitivity to update the parameters and iterate until a convergence criteria is met.  Note that the forward problem is non-linear and we solve this iteratively using Newton-Raphson iteration. We discretize both the forward and the inverse problem with a 3D brick element and a standard piecewise polynomial Lagrange basis function of degree 1. The parameters are updated using method of moving asymptotes (MMA) \cite{svanberg1987} until either the norm of the sensitivities or incremental objective reaches  $\varepsilon = 10^{-5}$. Algorithm \ref{alg:algo}  summarizes our approach.

The calculation of $\mu$ requires some care.   Since this term involves quantities from images that is typically pixelated, the integration is performed by locating pixels in each element and summing over these pixels (instead of quadratures).  Further, the imaging is done in the current configuration, and the current image $g(y,t)$ is in pixelated form with uniform pixel spacing in $y$ (current configuration).  However, we integrate it over the reference domain.  Therefore, we have to pull it back to the reference configuration by finding the values of $g(y(x,t),t)$ where $x$ is sampled uniformly in the reference domain.  We use bilinear interpolation to do so.  We compute $\nabla_y g$ by pixel differences in the current configuration and then interpolate to pull it back to the reference configuration.  While this term can be noisy, it is integrated against the shape function that is smooth on the scale of the pixels, and leading to a stable calculation of $\mu$.


\subsection{Contrast with the purely kinematic DIC}

Before we demonstrate the proposed method, we comment on the mathematical issues associated with the current practice of treating digital image correlation as a purely kinematic problem of finding deformations.  In this approach, we compare the image $g_0(x)$ of a decorated surface before deformation with that $g(y)$ after the deformation to obtain the deformation $y(x)$.  Specifically, we maximize the correlation between the reference and convected images over all possible deformations, or equivalently minimize the $L^2$ norm between the reference and convected images over all possible deformations,
\begin{equation} \label{eq:objdic}
    y  = \argmax \int_{\mathcal{S}}  g_0(x) g(y(x)) \, dA = \argmin  \int_{\mathcal{S}} \left| g_0(x) - g(y(x)) \right|^2 \, dA .
\end{equation}
The two formulations are equivalent since expanding the second integral gives us the first up to factor $(-2)$ and the $L^2$ norm of the images (that are independent of the deformation).

Unfortunately, this problem is ill-posed.  Note that the second problem, in (\ref{eq:objdic}) is the classical optimal transport problem of Monge (transporting a mass with density $g_0$ to a mass with density $g$), and $y$ is called the transport map.  Unfortunately, this problem is not mathematically well-posed, and solutions may not exist and minimizing sequences may not converge.  The relaxation, according to Kantorovich, is to look for minimizers over transport plans, $G(x,y)$, such that 
\begin{equation}
\min_G \int_{\mathcal{S}} G(x,y) \left| x - y \right|^2 \, dA \times dA
\end{equation}
subject to
\begin{equation}
\int_{\mathcal{S}} G(x,y) \, dA_y = g_0(x), \quad \int_{\mathcal{S}} G(x,y) \, dA_x = g(y).
\end{equation}
In essence, the density at some reference point $x$ can get spread over a region of the spatial domain, and the density at the spatial point $y$ may come from a region of the reference domain.  In short, the solution to the classical DIC problem (\ref{eq:objdic}) may not exist.

In practice, we make a finite dimensional ansatz on the deformation.  Two are common.  The first is local where we select a grid of points $\{\bar{x}^i\}$ and assume that the deformation is piece-wise affine in a sub-domain around these points
\begin{equation}
     y(x) = \sum_i (y^i + F^i(x-\bar{x}^i))\chi_i(x),
 \end{equation}
 where $\chi_i$ is the indicator function of the $i^\text{th}$ sub-domain around $\bar{x}^i$.  The second is non-local where the deformation is given in terms of a finite-dimensional (often finite-element based) basis set $\{\psi_i\}$,
 \begin{equation}
     y(x) = \sum_i y^i \psi_i(x)
 \end{equation}
These constraints yield a solution but there is no notion of convergence as the number of grid-points or basis functions goes to infinity.  Further, this constraint to finite dimensions makes the method fail or have poor accuracy when there are cracks, shear bands and shocks.

There are other difficulties with a purely kinematic approach.  We image a fixed region of space that may correspond to different parts of the body in different snapshots.  So, image registration limits us to small displacements.  One can stitch together multiple  snapshots, but errors can be compounded, and the method is inaccessible to dynamic events where one has very few snapshots.  Further, there are situations where the deformation and imaging interact as, for example, in the recent work of Lawlor {\it et al.} \cite{lawlor_2024} mentioned in the introduction.  

Note that we overcome these difficulties in our proposed approach since the deformation is obtained from the equilibrium equation (\ref{eq:gov}) and thus well-defined.
\section{Testing the idea with synthetic data} \label{sec:syn}

We test the idea by using synthetic data: we use a known constitutive relation, solve the forward problem with that constitutive relation and use it to generate synthetic data.  We then use the procedure described in Section \ref{sec:form} 
on the synthetic data to obtain the constitutive relation.  Comparing the inferred constitutive relation with the known constitutive relation provides us with a verification of the method.

\subsection{Homogeneous material}

\begin{figure}
\centering
\includegraphics[width=6.5in]{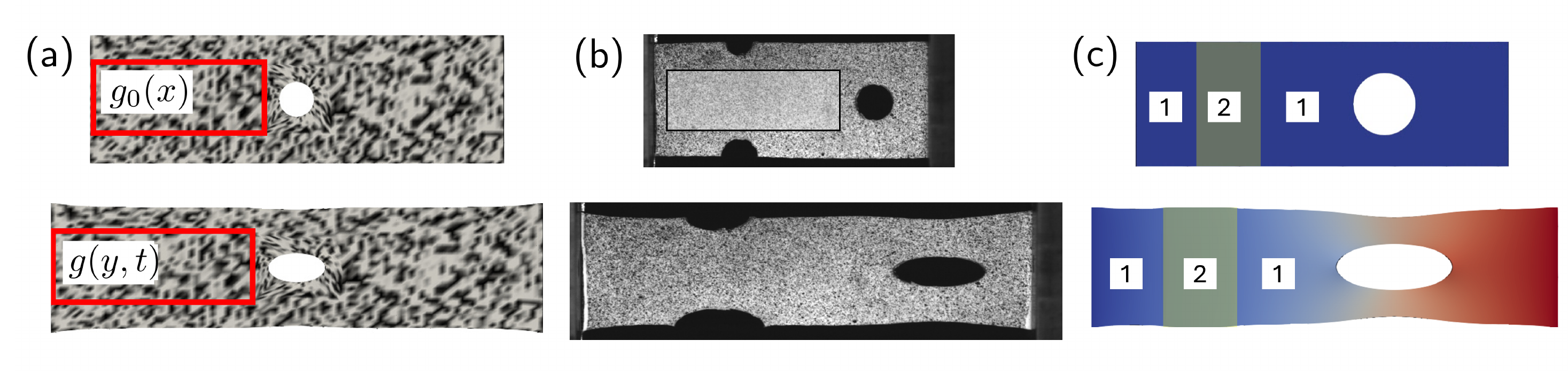}
\caption{Images of reference and deformed configuration for (a) testing, (b) experimental validation and (c) two material example. The region of interest to perform correlation is highlighted in red in (a). }
\label{fig:specimen}
\end{figure}

We consider a hyperelastic constitutive relation motivated by neo-Hookean:
\begin{equation}\label{eq:synNH}
    W=\frac{\mu}{2} \left( \text{tr}(F^TF)-3-2\log (\det F) \right) +\frac{\kappa}{2}(\det F -1)^2
\end{equation}
where the parameters are the shear and bulk moduli $P=\{\mu, \kappa\}$.  
We consider a specimen $(0,\ell) \times (0,w) \times (0,t)$ in the form of a rectangular strip with a central circular hole as shown in Figure \ref{fig:specimen}(a).  We create a speckle pattern ($g(x,0)$) from correlated solutions \cite{vic} also as shown.  We clamp this specimen with the constitutive relation (\ref{eq:synNH}) on the left ($y(\{0,x_2\})) = \{0,x_2\}$, and apply a horizontal displacement on the right $y(\{\ell,x_2\})) = \{ (1 + \dot{\bar{\varepsilon}}t) \ell, x_2 \}$ where the nominal strain rate $\dot{\bar{\varepsilon}}$ is taken to be extremely small.  The lateral surfaces, $x_2=0,w$, are traction-free.  We solve the equilibrium equation (\ref{eq:gov}) at each time $t$ to obtain the synthetic deformation $y(x,t)$.  We obtain the speckle pattern in the current configuration by convecting the reference image by the deformation:
\begin{equation}
g(y,t) = g_0(x(y,t))
\end{equation}
where $x(y,t)$ is the inverse deformation.  A snapshot is shown in Figure \ref{fig:specimen}(b).  We also compute the total reaction force $f(t)$ at $x_1=\ell$ at each instance.  We use  $\{g_0, \dot{\bar{\varepsilon}}\}$ and a number of snapshots of $\{g, f\}$ as our data.

\paragraph{Known constitutive relation.}

\begin{figure}
    \centering
    \includegraphics[width=5in]{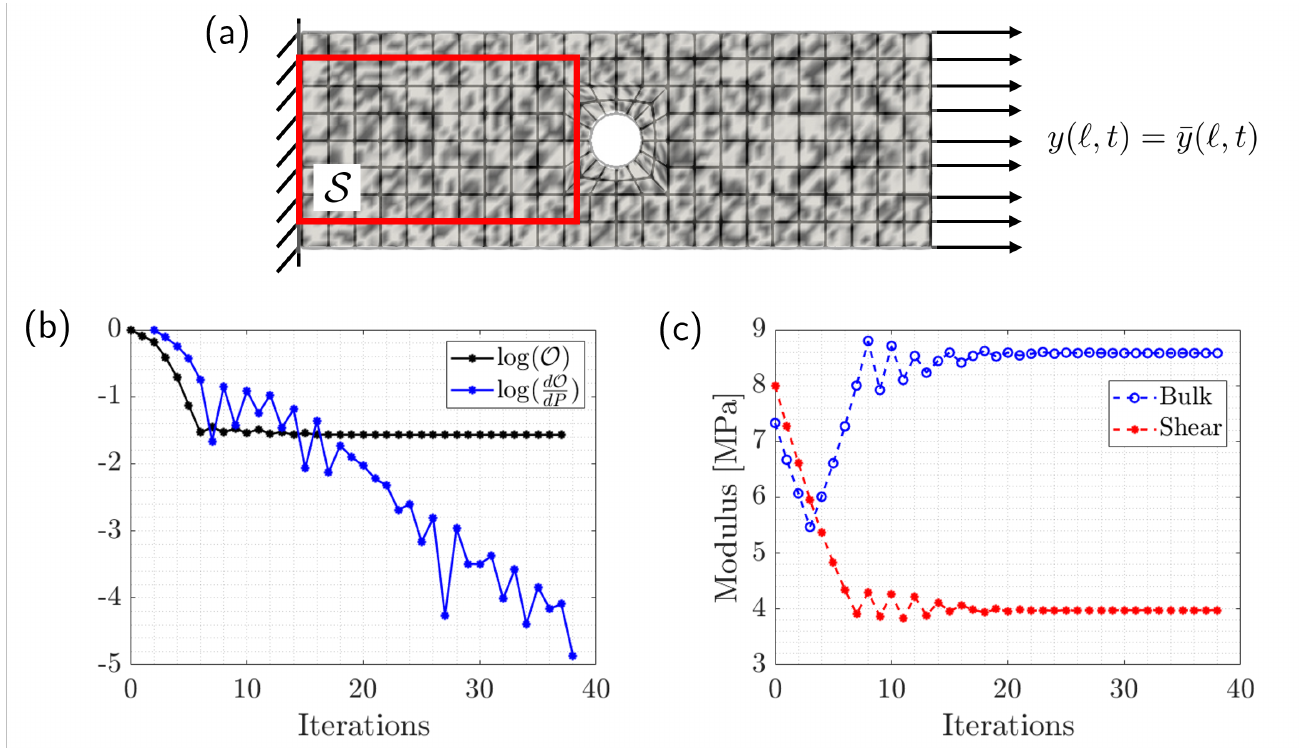} 
    \caption{Results for quasistatic uniaxial tensile test of specimen shown in (a). The red box is the region to perform image correlation, $\mathcal{S}$. (b) The normalized objective in (\ref{eq:obj}) and the combined sensitivity of material parameters with iterations. (c) The evolution of shear and bulk modulus. The synthetic data is generated with $\mu = 4\, \text{MPa}, \kappa = 8.6  \, \text{MPa}$ and the optimization is run with initial guess of $\mu = 8\, \text{MPa}, \kappa = 7.3  \, \text{MPa}$. }
    \label{fig:homo}
\end{figure}
A typical result is shown in Figure \ref{fig:homo}.  The data is generated with a initial parameters  $\mu = 4\, \text{MPa}, \kappa = 8.67  \, \text{MPa}$, and the calculation is initialized with $\mu = 8\, \text{MPa}, \kappa = 7.3  \, \text{MPa}$.  Figure \ref{fig:homo}(a) shows the reference image of the specimen with the finite element grid used for the simulations, the speckle pattern and the region of correlation $\mathcal{S}$ marked. The overall dimensions are $0.036 \times 0.012 \times 0.005$ m. Figure \ref{fig:homo}(b) shows the objective and the sensitivity, while Figure \ref{fig:homo}(c) shows the material parameter as the iteration proceeds.  We observe a quick drop in the objective, a steady drop in the sensitivity and a stable value for the parameters.  The recovered values are $\mu = 3.98\, \text{MPa}, \kappa = 8.58  \, \text{MPa}$, in close agreement with the values that generated the data.

\begin{table}
\centering
\begin{tabular}{|c|c|c|c|c|}
\hline
&\multicolumn{2}{|c|}{Initial guess}&\multicolumn{2}{|c|}{Recovered value} \\\hline
Case & Shear modulus, $\mu_i$ & Bulk modulus, $\kappa_i$ & Final shear, $\mu_f$ & Final bulk, $\kappa_f$ \\ \hline
\multicolumn{5}{|c|}{Set 1: Generate data using $\mu =$ 4 MPa, $\kappa = $ 8.67 MPa}  \\ \hline 
1a    & 4 MPa   & 38 MPa  & 3.95 MPa   & 8.63 MPa \\ \hline
1b    & 1 MPa   & 9.67 MPa   & 4.03 MPa   & 8.35 MPa \\ \hline
1c    & 0.4 MPa   & 3.8 MPa   & 3.95 MPa   & 8.61 MPa \\ \hline
1d    & 40 MPa   & 387 MPa   & 3.95 MPa   & 8.50 MPa \\ \hline
1e    & 0.4 MPa   & 0.36 MPa   & 3.98 MPa   & 8.52 MPa \\ \hline
1f    & 8 MPa   & 7.34 MPa   & 3.98 MPa   & 8.58 MPa \\ \hline
1g    & 5 MPa   & 6.67 MPa   & 3.99 MPa   & 8.61 MPa \\ \hline
\multicolumn{5}{|c|}{Set 2: Generate data using $\mu =$ 4 MPa, $\kappa = $ 38.6 MPa}  \\ \hline 
2a    & 1 MPa   & 3 MPa  & 3.99 MPa   & 33.3 MPa \\ \hline
2b    & 8 MPa   & 10 MPa   & 3.99 MPa   & 33.3 MPa \\ \hline
2c    & 2.2 MPa   & 7 MPa   & 4.38 MPa   & 36.0 MPa \\ \hline
\multicolumn{5}{|c|}{Set 3: Generate data using $\mu =$ 0.5 MPa, $\kappa = $ 4.83 MPa}  \\ \hline 
3a    & 1 MPa   & 2 MPa  & 0.55 MPa   & 4.48 MPa \\ \hline
3b    & 0.1 MPa   & 0.8 MPa   & 0.49 MPa   & 4.19 MPa \\ \hline
3c    & 0.3 MPa   & 1.0 MPa   & 0.5 MPa   & 4.16 MPa \\ \hline 
\end{tabular}
\caption{Performance with various initial guesses and various sets of synthetic data.}
 \label{table: IG_NH}
\end{table}

We repeat the simulation with several values of the initial guess, and data generated with three sets of parameters.  The results are shown in Table \ref{table: IG_NH}.  We see that we get very good recovery in each case.   

\begin{figure}
    \centering
    \includegraphics[width=6.5in]{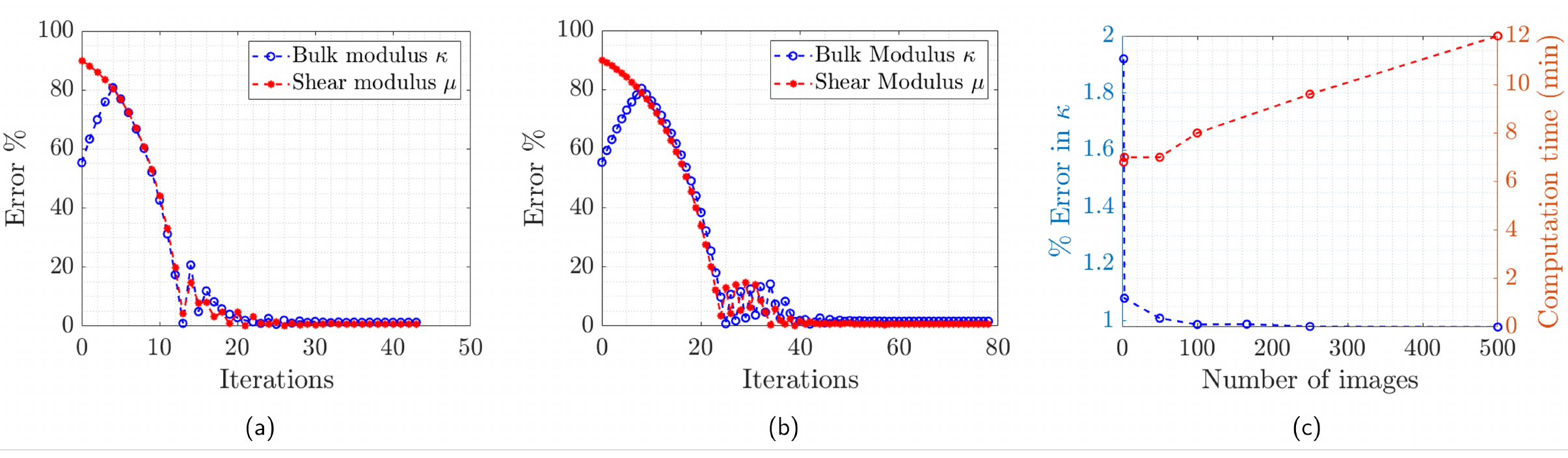}
    \caption{Performance with (a) 10\% random noise in for synthetic images with 720x240 pixels and (b) 360x120 pixels. (c) various number of images for inversion.}
    \label{fig:noise}
\end{figure} 

We now turn to noisy data.   Images obtained from experiment has noise depending on the experimental setup and pixel accuracy, and so we seek to test our approach against noisy data.  We generate images with 720x240 pixels and 360x120 pixels within the domain and add 10\% uniformly distributed random noise to the speckle values for all pixels. The range of speckle values is between 0 and 1, hence the maximum absolute noise value is 0.1.  As before, the synthetic data is generated with $\mu = 4 \,\text{MPa}$ and $\kappa = 8.67 \, \text{MPa}$. The initial guess of parameters is $\mu = 0.4 \,\text{MPa}$ and $\kappa = 0.44 \, \text{MPa}$. The evolution of material parameters can be seen in Figure \ref{fig:noise} (a,b). We note that the method is able to recover both parameters with just 4\% error. 

We study the performance of our method with 2, 3, 50, 100, 250, and 500 snapshots of $g(y,t)$ resulting in errors of  $1.9\%, 1.1\%, 1.0\%, 0.99\%, 0.98\%, 0.97\%$ in the bulk modulus respectively. The error is bulk modulus and computational time per iterations can be seen in Figure \ref{fig:noise}(c). In all cases, the number of optimization iterations using MMA method is 35, 35, 37, 39, 39, and 34, respectively. Thus, we get reasonable recovery with two snapshots and this improves further with more snapshots. 

\begin{table}
\centering
\begin{tabular}{|c|c|c|c|c|c|}
\hline
&\multicolumn{3}{|c|}{Total pixels}&\multicolumn{2}{|c|}{Mesh refinement} \\\hline
$\%$ Error & 180 $\times$ 60 & 360 $\times$ 120  & 720 $\times$ 240 & 1$\times$ mesh & 2$\times$ mesh \\ \hline 
Shear modulus, $\mu$       & 0.5\% & 0.4\%   & 0.4\%  & 0.8\%  & 0.4\% \\ \hline
Bulk modulus, $\kappa$        & 2.4\%    & 1.3\%  &  0.6\%    & 1.5\%   & 1.3\% \\ \hline
\end{tabular}
\caption{Performance tests: various pixel per mesh and mesh refinement.}
 \label{table: compstudy}
\end{table}

We conclude this study by looking at a few computational issues.   We study the performance of the method under refinement for both the image and the finite element discretization.  In each case, the synthetic data is developed using  parameters $\{ \mu, \kappa \} = 4 \text{ MPa}, 8.67 \text{ MPa}$, and the simulations are initialized with parameters $\{ \mu, \kappa \} = 8 \text{ MPa}, 7.34 \text{ MPa}$. Table \ref{table: compstudy} summarizes the effect of refining the image (pixel resolution) while keeping the finite element discretization fixed. We see that the errors are still small, though it degrades a little with decreasing image resolution. The effect of fixing the image resolution (at 360x120 pixels) and refining the mesh can also be seen in Table \ref{table: compstudy}. We find slight improvement in the recovery with a refined mesh. \\



\vspace{0.1in}

Putting all these together, we conclude that the proposed method successfully recovers the parameters of a known constitutive relation.

\paragraph{Unknown constitutive relation.}

We apply our approach to synthetic data generated as described above with (\ref{eq:synNH}).  However, in contrast to the previous example, we do not assume a knowledge of the form of the constitutive relation.  Instead, we assume a different parametrized form, a compressible Mooney-Rivlin relation
\begin{equation}
    W = C_1(I_1-3) + C_2(I_2-3) + C_3(I_1-3)^3 + \kappa/2(I_3-1)^2,
\end{equation}
with parameters $P=\{C_1, C_2, C_3, \kappa\}$  where $I_1, I_2, I_3$ are the principal invariants of the left Cauchy-Green tensor.   We proceed as before and obtain the parameters $C_1 = 2.1$ MPa, $C_2 = 0.001$ MPa, $C_3 = 0.14$ MPa, and $\kappa = 12$ MPa.

\begin{figure} 
    \centering
    \includegraphics[width=5.0in]{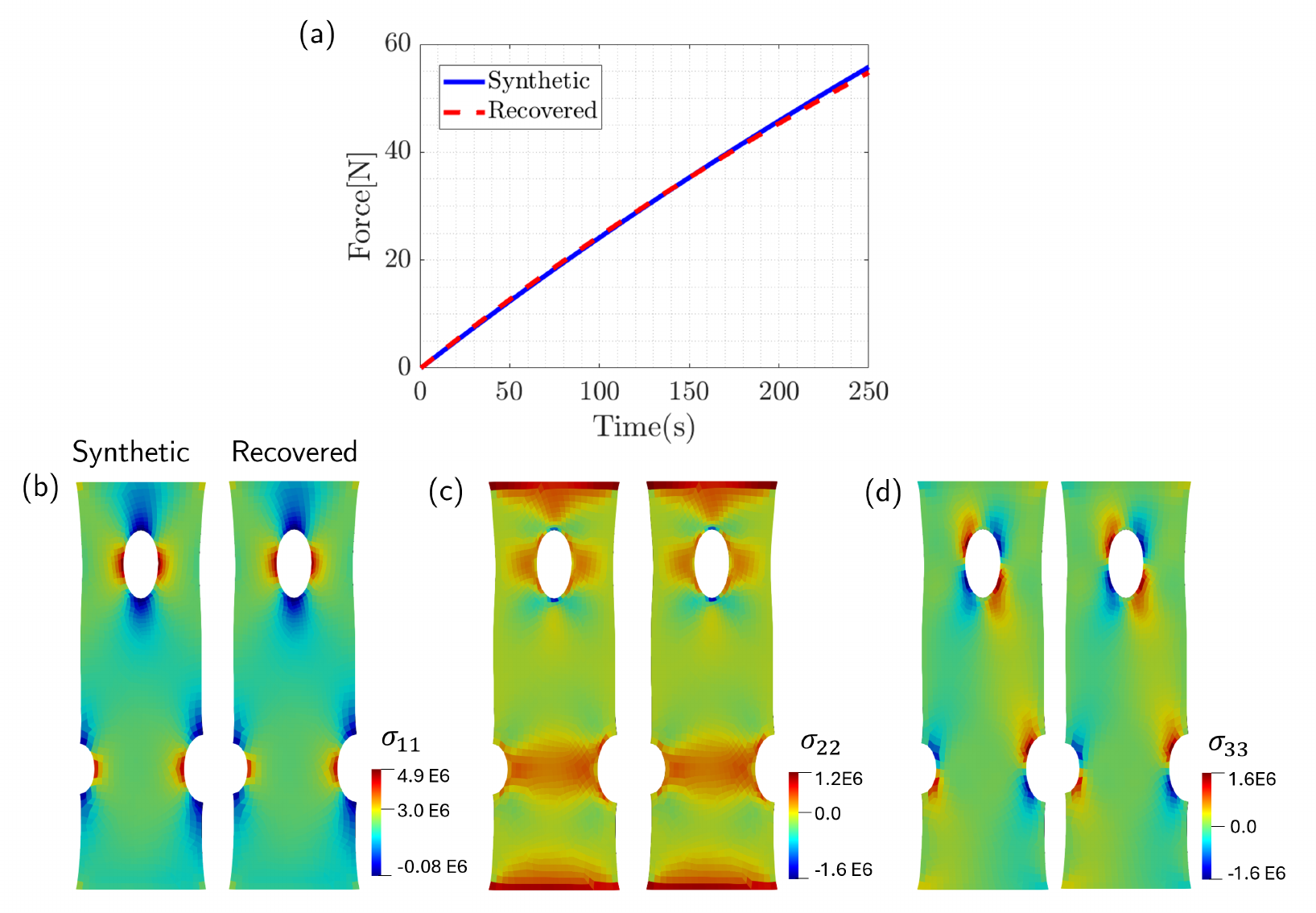}
    \caption{Comparison of the synthetic data and recovered constitutive relation using an independent test. (a) Force and (b-d) stress for the verification specimen at a nominal strain of 0.25.}
    \label{fig:unknown}
\end{figure} 

To verify that the recovered constitutive relation still captures the actual constitutive behavior, we perform an independent simulation on a validation specimen with a different geometry as shown in Figure \ref{fig:specimen} (overall dimensions $0.05 \times 0.02 \times 0.005$ m). Figure \ref{fig:unknown}(a) compares the measured and computed force between synthetic and recovered constitutive relations. Figure \ref{fig:unknown}(b-d) shows the comparison of $\sigma_{11}$, $\sigma_{22}$, and  $\sigma_{22}$ at a nominal strain of $25\%$.
We define the overall total error in the stress and strain fields to be
\begin{equation}
    e_\sigma = \frac{\int_0^t\int_\Omega (\sigma_{synthetic} - \sigma_{R}(P))^2 \,d\Omega \,dt }{\int_0^t\int_\Omega (\sigma_{synthetic})^2 \,d\Omega \,dt }, \quad  e_\epsilon = \frac{\int_0^t\int_\Omega (\epsilon_{synthetic} - \epsilon_{R}(P))^2 \,d\Omega \,dt }{\int_0^t\int_\Omega (\epsilon_{synthetic})^2 \,d\Omega \,dt },
\end{equation}
and find 0.022\% error in stress and 0.025\% error in strain. 
\vspace{0.1in}

Therefore, we conclude that the proposed method successfully recovers the constitutive relation even without {\it a priori} knowledge of the form of the constitutive relation.

\subsection{Heterogeneous material}

The method above can also be used when the material is piecewise uniform, as long as the domains of uniformity are identified {\it a priori}.  The derivation of the sensitivity using the adjoint equation described earlier can easily be generalized to this case.  The adjoint equation and the sensitivity are exactly as in (\ref{eq:adj}) and (\ref{eq:sen}) respectively, except $W$ depends on $x$.   Importantly, the adjoint problem is solved only once per iteration, and the parameters for all the materials are updated simultaneously from the resulting sensitivity.

We now demonstrate the performance of our method when we have two materials as shown in Figure  \ref{fig:specimen}(c).  We assume the constitutive law (\ref{eq:synNH}) for each material, but with different parameters.  We generate synthetic data with $(\mu_1,\kappa_1) = (4 \, \text{MPa}, 8.67 \, \text{MPa})$ and $(\mu_2,\kappa_2) = (20 \, \text{MPa}, 136 \, \text{MPa})$ as described above.  The results of our algorithm for various initial guesses are shown in Table \ref{table: IG_NH_BI}.  We observe excellent recovery as before.

\begin{table}
\centering
\begin{tabular}{|c|c|c|c|c|}
\hline
&\multicolumn{2}{|c|}{Initial guess}&\multicolumn{2}{|c|}{Recovered value} \\\hline
Case & Shear modulus, $\mu_i$ & Bulk modulus, $\kappa_i$ & Final shear, $\mu_f$ & Final bulk, $\kappa_f$ \\ \hline
\multicolumn{5}{|c|}{Material 1: Generate data using $\mu =$ 4 MPa, $\kappa = $ 8.67 MPa}  \\ \hline 
1       & 1 MPa   & 2.17 MPa  & 4.03 MPa   & 8.42 MPa \\ \hline
2       & 20 MPa   & 9.34 MPa &  4.01 MPa   & 8.28 MPa \\ \hline
3       & 8 MPa & 7.34 MPa   & 4.16 MPa   & 8.44 MPa \\ \hline
4       & 1 MPa  & 1.67 MPa  & 3.95 MPa & 8.59 MPa \\ \hline
5       & 0.8 MPa   & 7.73 MPa   & 3.96 MPa   & 8.09 MPa  \\ \hline
6       & 8 MPa   & 3.73 MPa & 4.28 MPa   & 8.06 MPa \\ \hline
\multicolumn{5}{|c|}{Material 2: Generate data using $\mu =$ 20 MPa, $\kappa = $ 136 MPa}  \\ \hline 
1	& 15 MPa   & 45 MPa & 19.8 MPa   & 127 MPa \\ \hline
2	& 40 MPa   & 36.7 MPa & 20.1 MPa   & 119 MPa \\ \hline
3	& 15 MPa   & 145 MPa & 20.1 MPa   & 135 MPa \\ \hline
4	& 40 MPa   & 187 MPa  & 20 MPa   & 130 MPa \\ \hline
5	& 10 MPa   & 96.7 MPa & 19.7 MPa   & 145 MPa \\ \hline
6	& 25.0 MPa   & 242 MPa & 22.4 MPa   & 201 MPa \\ \hline
\end{tabular}
\caption{Performance in the two material situation and various initial guesses.}
 \label{table: IG_NH_BI}
\end{table}

\section{Experimental demonstration} \label{sec:exp}

We now apply the proposed method on experimental data obtained using a natural rubber specimen (McMaster-Carr, 87145k411).  We consider a  specimen of size $ 0.036 \times 0.012 \times 0.0015$ m with a hole in the center as shown in Figure \ref{fig:specimen}(a) and apply a speckle pattern of similar size to that in the figure. We clamp the left end, and apply a uniaxial displacement to the right at a nominal strain rate of $\dot{\epsilon} = 0.001/$ up to a nominal strain of 45\% on the Instron E3000.  As the sample deforms, we record the boundary force and image the sample at 2 frames per second, with resolution of 1024 by 1024 pixels of the field of view (Photron Fastcam NOVA S12 camera and Tokina AT-X PRO lens, 100F 2.8D).   We use bilinear interpolation of the image to downsample the images to  to $720$px $\times 240$px.

We then perform our method with the series of images collected. We assume that our material is well-described by a compressible Mooney-Rivlin constitutive law, 
\begin{equation}
    W = C_1(I_1-3) + C_2(I_2-3) + \kappa/2(I_3-1)^2,
\end{equation}
with parameters $P=\{C_1, C_2, \kappa\}$  where $I_1, I_2, I_3$ are the principal invariants of the left Cauchy-Green tensor. We obtain $C_1 = 0.33$ MPa, $C_2 = 0.21$ MPa, and $\kappa = 7.53$ MPa as the parameters from the experiment.

\begin{figure} [t]
    \centering
    \includegraphics[width=6.0in]{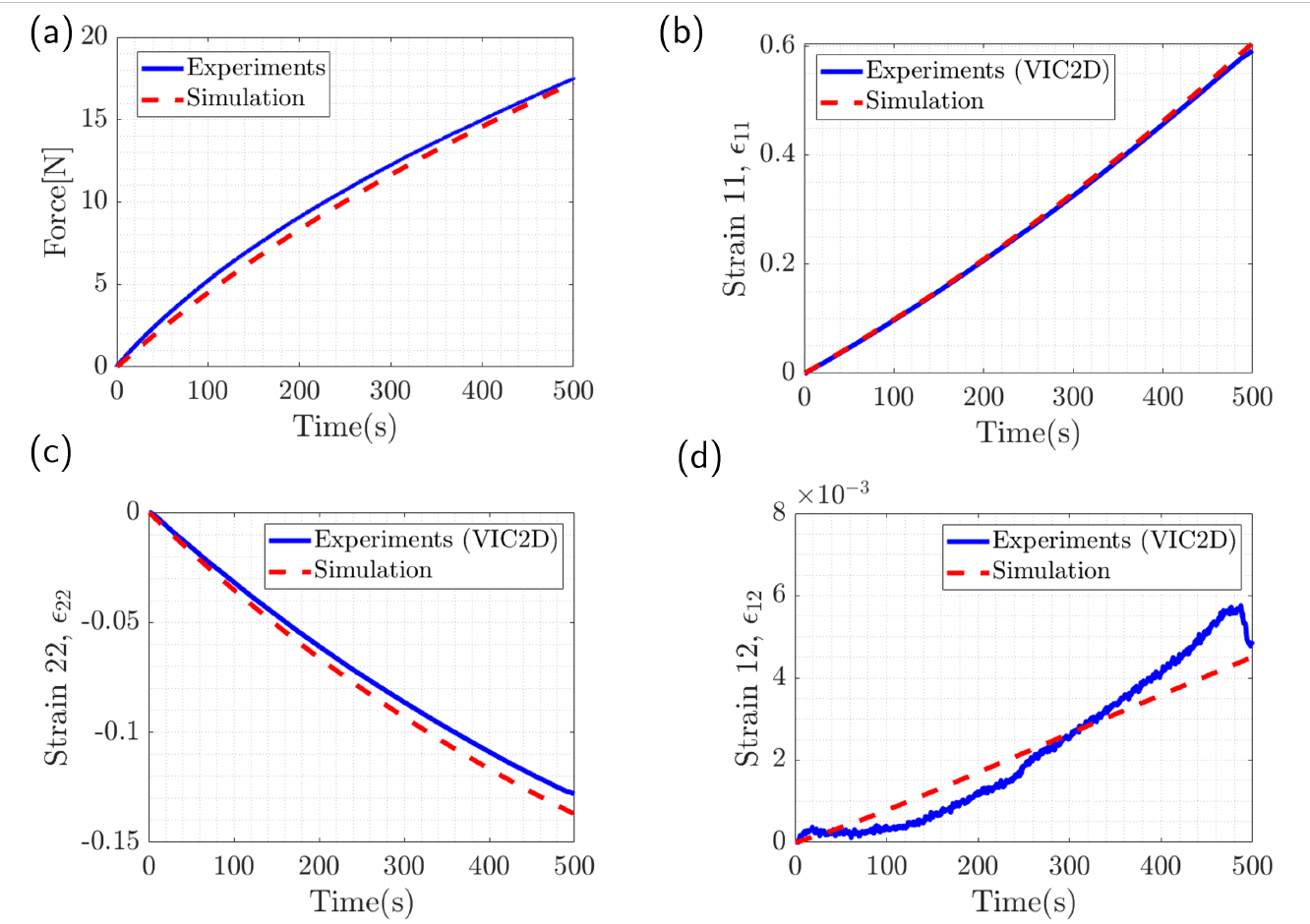}
    \caption{Comparison of the measured and computed (a) force and (b-d) average strain response for the verification specimen.  The computed response uses the recovered parameters and the strains are averaged over a region marked in Figure \ref{fig:specimen} in using the recovered parameters for the verification specimen.
    }
    \label{fig:avecomp}
\end{figure} 

To verify that these recovered constitutive behavior accurately describes the material, we conduct an independent experimental test on specimen with a different geometry shown in Figure \ref{fig:specimen} (overall dimensions $0.05 \times 0.02 \times 0.0015$ m). As before, the sample is deformed at $\dot{\epsilon} = 0.001/$s up to 50\% strain, and we record the force and image the specimen every 2 fps.      Figure \ref{fig:avecomp}(a) compares the measured and computed force response using a finite element calculation and the recovered parameters.  We find an excellent match. 

\begin{figure}
    \centering
    \includegraphics[width=6.0in]{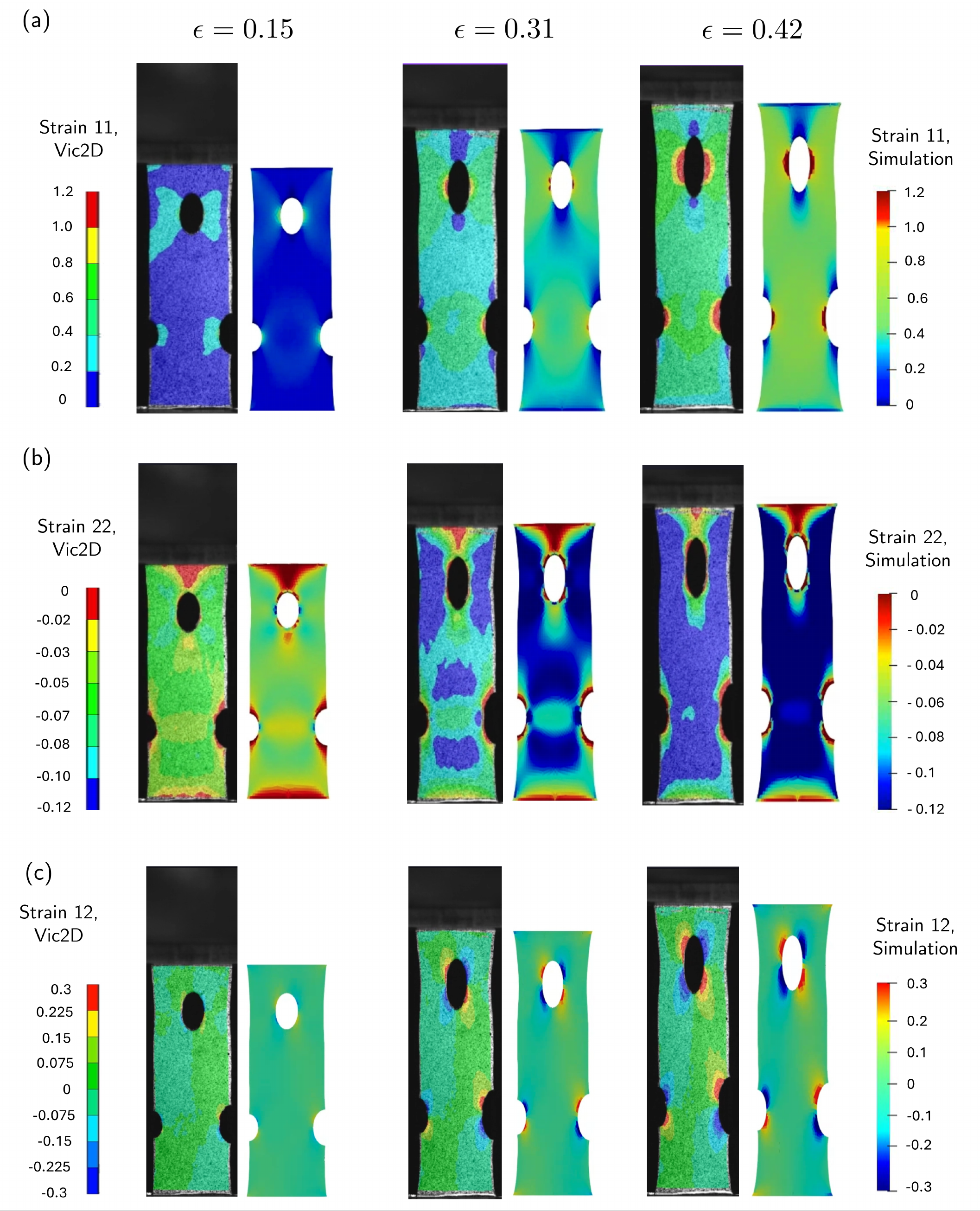}
    \caption{Experimental images used for verification of material parameters of a natural rubber specimen.}
    \label{fig:straincomp}
\end{figure} 

We then use the recorded images and the commercial VIC2D software (Correlated Solutions, Columbia, SC) to obtain the (Lagrangian) strain field.   Figure \ref{fig:straincomp} compares the strain field obtained from VIC2D with that computed using finite element analysis and the recorded parameters at three loading stages.  Figure \ref{fig:avecomp}(b-d) compares the strains averaged over a region marked in Figure \ref{fig:specimen}.   Again we find an excellent match.
We note that the discrepancy in $\epsilon_{12}$ is comparable to the strain uncertainty of VIC-2D. The uncertainty is determined by measuring noise floor of strain values in VIC2D by taking a series of five still images of a control sample. Under no deformation, the standard deviation of $\epsilon_{11},\epsilon_{22},\epsilon_{12}$ are 0.0006, 0.0005, and 0.0004 respectively.
\vspace{0.1in}

We conclude that the proposed method successfully recovers the constitutive relation from an experimental test.

\section{Conclusion} \label{sec:conc}

We address the problem of learning constitutive relations of materials from experiments.  We seek to use the power of full field observation techniques like digital image correlation.  In a departure from previous work that breaks down the inverse problem into two nested optimization problem, first obtaining deformation from images and second obtaining constitutive relations from deformation and force, we propose an integrated approach.  Specifically, we formulate the problem of obtaining constitutive relations from raw data (images and force) as partial differential equation constrained optimization problem, and solve it using the adjoint method.  The integrated approach overcomes many of the problems associated treating digital image correlation as a purely kinematic problem of obtaining deformation from images, as well as the problems associated with a multi-level optimization.  We demonstrate this approach on finite elasticity using both synthetic and experimental data.  We view this as a first step in a larger program, and plan to address history dependent behavior, shocks, fracture and other phenomena in future work.

\subsection*{Acknowledgement}  We are grateful for the financial support of the National Science Foundation (2009289) and the Army Research Office (W911NF-22-1-0269).


\end{document}